\documentclass[fleqn,10pt]{wlscirep}
\usepackage[utf8]{inputenc}
\usepackage[T1]{fontenc}

\newcommand{\ket}[1]{|#1\rangle}

\title{A versatile laser-machined rf trap for arrays of 100+ ions}

\author[1]{Frank G. Schroer}
\author[1]{Ilyoung Jung}
\author[1]{Thomas W. Burkle}
\author[1]{Jack B. Lyons}
\author[1]{Joseph W. Van Vlack}
\author[1]{Joseph Ezuma}
\author[1,2,*]{Philip Richerme}

\affil[1]{Indiana University Department of Physics, Bloomington, Indiana 47405, USA}
\affil[2]{Indiana University Quantum Science and Engineering Center, Bloomington, Indiana 47405, USA}

\affil[*]{richerme@iu.edu}

\begin{abstract}
Large ion crystals in diverse geometries are a key resource for quantum simulation experiments. In this work, we introduce a macroscopic rf trap that supports a wide variety of one-dimensional ion configurations as well as lateral two-dimensional crystals with more than 100 ions. Our design is based on precision-machined fused silica wafers that are stacked to form the trap structure. Ten independently biased electrodes provide flexible control over the axial potential, enabling long one-dimensional crystals, isospaced ion strings, split-well chains, and two-dimensional arrays with tunable aspect ratios. We present the design and fabrication process for this trap and demonstrate the ability to tune the radial secular frequencies, detect and compensate micromotion, rotate the principal axes, and characterize trapped ion heating rates. All trap design and documentation files are freely available alongside this work, to facilitate adoption and further development within the ion trap community.

\end{abstract}
\begin{document}

\flushbottom
\maketitle

\thispagestyle{empty}

\section*{Introduction}
The rf Paul trap has served for over half a century as a means to confine charged particles for precision measurements, atomic clocks, and quantum information processing \cite{paul1990electromagnetic,ludlow2015optical,wineland1998experimental,monroe2021programmable}. Over this long history, the original hyperbolic ring trap geometry has evolved in countless ways to meet the scientific demands of specialized experiments \cite{siverns2017ion}. For example, traps advancing towards scalable quantum computation often employ sophisticated engineering and manufacturing processes to produce devices with hundreds of electrodes \cite{maunz2016high,revelle2020phoenix,pino2021demonstration,moses2023race,chen2024benchmarking}, integrated optics \cite{streed2011imaging,true2011demonstration,mehta2020integrated,ivory2021integrated,corsetti2026integrated}, or integrated microwave circuitry \cite{ospelkaus2011microwave,allcock2013microfabricated,shappert2013spatially,lekitsch2017blueprint,hughes2025trapped,bowers2025robust}. Unfortunately, the design and implementation of such complex traps are often inaccessible to many groups due to their long development times and the high costs of fabrication and testing.

For quantum simulation experiments, where fast shuttling and discrete two-qubit gate operations are generally not required \cite{monroe2021programmable}, conventional macroscopic traps remain a cornerstone technology for confining ions quickly and at low expense. Example geometries include linear rod traps \cite{drewsen1998large,olmschenk2007manipulation,shapira2025programmable}, blade-style traps \cite{gulde2003universitat,friedenauer2008simulating,mcloughlin2011versatile,ulm2013observation,jurcevic2014quasiparticle,debnath2016demonstration,xie2021open}, and cylindrical traps \cite{bergquist1986observation,yoshimura2015creation,huntemann2016single,ivory2020paul,pashinsky2025structural}, which all enjoy relatively straightforward assembly and the reliability associated with deep confining potentials. However, the tradeoffs for this simplicity are typically low secular frequencies (which complicate ion cooling and can reduce experimental fidelities) and limited degrees of freedom with which to shape trapping potentials.

Here, we introduce an rf Paul trap for quantum simulation experiments designed to support large ion crystals in a variety of 1D and 2D configurations. Our trap, which is composed of laser-machined stacked wafers \cite{madsen2006advanced,pyka2014high,brewer2019al+,ragg2019segmented,jordan2025scalable}, leverages the tight tolerances achievable with precision glass machining while avoiding the complexity of fully monolithic traps \cite{wang2020coherently, kiesenhofer2023controlling,guo2024site,menon2026monolithic}. Our design is versatile: 20 dc electrodes (10 grounded, 10 with independent biases) and carefully chosen electrode dimensions enable tailored trapping potentials for long 1D chains, equispaced ion strings, ions in multiple wells, and lateral 2D crystals containing more than 100 ions. Furthermore, our design offers access to this wide feature set while remaining easy to fabricate and assemble. A full list of parts, assembly instructions, and CAD models for this trap are freely available in tandem with this publication \cite{schroer2026data}.

The trap design described in this work offers several advantages for quantum simulation experiments compared to prior macroscopic traps. First, the higher precision achievable with laser micromachining enables smaller trap feature sizes and ion-electrode distances, leading to larger trap secular frequencies and improved ion cooling. Second, higher precision fabrication leads to fewer trap asymmetries and misalignments compared to hand-assembled versions. Though this is typically compensatable for 1D chains, unwanted couplings between axial and radial motion in 2D crystals can pose severe limitations on ion number due to rapid rf-driven heating effects \cite{d2021radial,xie2021open}. Finally, compared with most linear or blade-style rf traps, laser micromachining offers the opportunity to include many more dc electrodes. This capability allows for multiple trap potential configurations and the confinement of ions in lattice structures inaccessible to simpler macroscopic traps.  

\section*{Results}

\subsection*{Trap Design and Modeling}
\begin{figure}[t]
    \centering
    \includegraphics[width=\columnwidth]{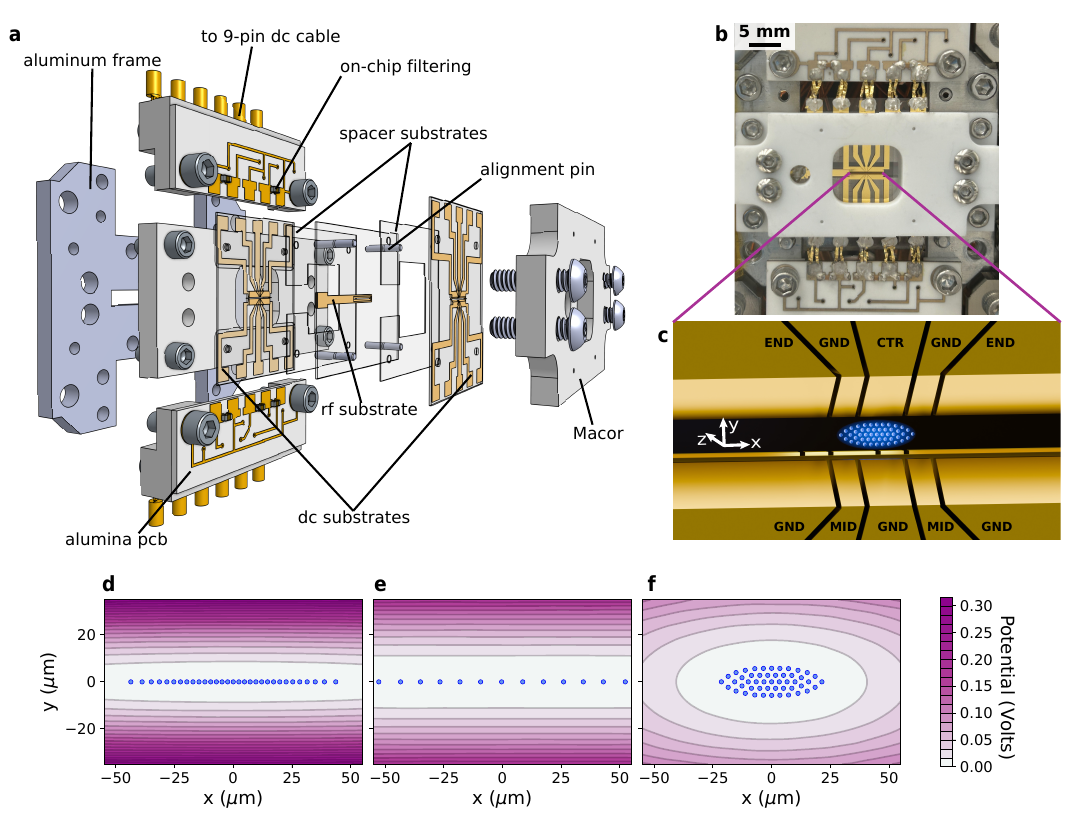}
    \caption{(a) Exploded CAD assembly showing the trap substrates, mounting system, and electrical connections. (b) Picture of the mounted substrates and electrical connections; (c) Schematic of the central trapping region, showing the electrode geometry and coordinate axes. Voltages are applied to endcap (END), midcap (MID), and center (CTR) electrodes, which alternate with grounded (GND) electrodes. Simulated electrical potentials in the trapping region predict the ability to confine ions in multiple geometries, including (d) long 1D chains, (e) uniformly-spaced 1D strings, and (f) lateral 2D crystals.}
    \label{fig:trapcad}
\end{figure}

The trap presented here is designed to support long 1D chains, equispaced 1D crystals, ions split between two wells, and large lateral 2D arrays. These requirements impose stringent demands on the shape and stiffness of the axial trap potential, which may be realized by controlling voltages applied to multiple dc electrodes at the $\sim 100-300~\mu$m scale. However, such characteristic dimensions and tight tolerances are challenging to achieve with standard machining approaches such as end milling or wire electrical-discharge machining. To meet these requirements, we pursue a trap implementation assembled from fused-silica substrates fabricated via selective laser etching (SLE). This ensures that the multiple sub-millimeter scale electrodes on each substrate are co-aligned to within the 1-3$~\mu$m tolerances of the laser machining process. 

The stacked wafer trap shown in Fig.~\ref{fig:trapcad}(a-c) consists of five layers: two outer dc electrode layers ($250~\mu$m thick), one central rf layer ($250~\mu$m thick), and two spacing layers ($125~\mu$m thick). Each layer contains four 1.000 mm alignment holes, into which $1.000^{+ 0.000}_{- 0.014}$~mm dowel pins are inserted before the stack is fastened together. Misalignments of the trap wafer stack are therefore constrained to the same level as SLE tolerances. Finite-element simulations (described in more detail below) predict only micron-scale deviations in the trap potential minimum due to these machining and assembly tolerances, which are easily compensated with applied trap voltages.

We used finite element method (FEM) simulations to guide and optimize the specific choices of trap electrode number, geometry, and bias voltages. We found that 20 dc electrodes (10 with independent biases, and 10 permanently grounded) offered sufficient control over the trapping potential to achieve all desired crystal geometries. Of the independent electrodes, the four outermost segments (`endcaps') are 2.44~mm long to provide overall axial confinement; four additional electrodes (`midcaps') are $110~\mu$m long and located axially inwards to allow for enhanced potential tunability; two center electrodes are $200~\mu$m long and provide additional axial and stray field compensation control. Our simulations indicate the importance of keeping the center and midcap electrode lengths short, so that lateral 2D crystals can be produced without requiring hundreds of volts applied to the endcaps. Similarly, simulations show that the midcaps allow for improved tuning of anharmonic coefficients in the axial potential, which is critical for equispaced and split-well chains. In addition, they facilitate rotation of the principal axes, which is important for minimizing micromotion in lateral 2D crystals. 

The distance between the central trap axis and mid-layer rf electrode is $100~\mu$m, which allows for relatively high radial secular frequencies and improved ion cooling compared with traditional rod or blade traps. However, this choice of rf electrode dimension will ultimately limit the number of ions that can be confined in lateral 2D arrays. Assuming that ions are kept at least $50~\mu$m away from the nearest trap surface to minimize ion heating, we estimate that $N\approx 500$ ions may be confined in a symmetric lateral 2D crystal, and $N\approx 1000$ in a 2D crystal with large aspect ratio. In practice, we expect more significant limitations on ion number due to off-axis micromotion and collisions with background gas in our room temperature apparatus.

We use our FEM simulations to predict the 1D and 2D ion crystal configurations that result from applying various voltage sets to our trap electrodes. First, we numerically calculate the dc potential and rf pseudopotential contributions to find the total potential near the center of the trap. We then determine the equilibrium configuration for a crystal of $N$ ions, each with mass $m$ and charge $Q$, by finding the ion coordinates that minimize the potential energy \cite{wang2015quantum}

\begin{equation}
\label{eq:ionPE}
        V(x,y,z) = \sum_{i=1}^{N}\frac{1}{2}m[f_{\text{axial}}(x_i)+\omega_{y}^2 y_i^2 + \omega_z^2z_i^2] +  \frac{Q^2}{4\pi\epsilon_o}\sum_{i < j}^N \frac{1}{\sqrt{(x_i-x_j)^2 + (y_i-y_j)^2 + (z_i-z_j)^2}}.
\end{equation}
In Eq. \ref{eq:ionPE}, the first term represents the potential due to the trapping fields, and the second term accounts for the Coulomb energy between each pair of ions. The function $f_{\text{axial}}(x)$ describes the axial potential (in units of m$^2$/s$^2$), which is not assumed to be harmonic. In fact, significantly anharmonic potentials are required to produce equispaced or split-well 1D chains and are implementable using our 10 independent dc electrodes. 

Predicted equilibrium ion geometries are shown in Fig.~\ref{fig:trapcad}(d-f) for three different voltage configurations. In Fig.~\ref{fig:trapcad}(d), the confinement along the axial direction is made very weak compared to the radial potential and results in a long 1D chain. By systematically raising the potential of the central electrode, the center ions can be pushed further apart to generate a chain with nearly uniform spacing (Fig.~\ref{fig:trapcad}(e)). Conversely, an equilibrium 2D configuration in the lateral ($xy$) direction can be produced by increasing the endcap voltages, such that the axial and $y-$radial confinements become comparable (Fig.~\ref{fig:trapcad}(f)). Our FEM simulations predict a robust set of voltage parameters for achieving each of these ion geometries, while staying far away from structural phase transitions of the Coulomb crystal \cite{d2021radial}.

\subsection*{Trap Characterization}
Near a local minimum, the time-dependent potential in an rf trap may be approximated as \cite{wineland1998experimental}
\begin{equation}
\label{eq:trappotential}
\Phi(\vec{r},t)=\Phi_\text{rf}(\vec{r},t)+\Phi_\text{dc}(\vec{r}) =
    \frac{V_0 \cos(\Omega_t t)}{2d_0^2}(y^2-z^2) + \frac{\kappa U_0}{2x_0^2}(2x^2-\gamma y^2-\zeta z^2)
\end{equation}
where $V_0$ is the rf voltage amplitude, $\Omega_t$ is the rf frequency, and $d_0$ and $x_0$ are the radial and axial trap dimensions.  In Eq. \ref{eq:trappotential}, $\kappa$ is a factor of order unity that scales the applied dc voltage $U_0$ based on the electrode geometry. We have also introduced geometric factors $\gamma$ and $\zeta$ to account for the radial asymmetry of the trap, with $\gamma+\zeta=2$ needed to satisfy Laplace's equation.

Given the form of the trap potential in Eq. \ref{eq:trappotential}, we use the pseudopotential approximation to estimate the radial trap frequencies
\begin{equation}
    \label{eq:trapfreqs}
    \omega_y=\sqrt{\frac{Q}{m}\left(\frac{QV_0^2}{2md_0^4\Omega_t^2}-\frac{\kappa\gamma U_0}{x_0^2}\right)}~~~~~;~~~~~
    \omega_z=\sqrt{\frac{Q}{m}\left(\frac{QV_0^2}{2md_0^4\Omega_t^2}-\frac{\kappa\zeta U_0}{x_0^2}\right)}
\end{equation}
where the frequency splitting between the radial modes is tunable based on the geometric factors $\gamma$ and $\zeta$. Under typical experimental conditions, we set $\omega_z > \omega_y$ with a frequency splitting of at least $2\pi\times 100$~kHz so that the modes in each direction are clearly separated. For the various 1D ion configurations, the axial potential is shaped using the 10 dc electrodes and is weak relative to the radial confinement. For lateral 2D crystals, we increase $\kappa U_0$ and tune $\gamma > \zeta$, such that the axial and radial$-y$ confinements are weak compared to the radial$-z$ confinement.

\begin{figure}[t]
    \centering
    \includegraphics[width=.7\columnwidth]{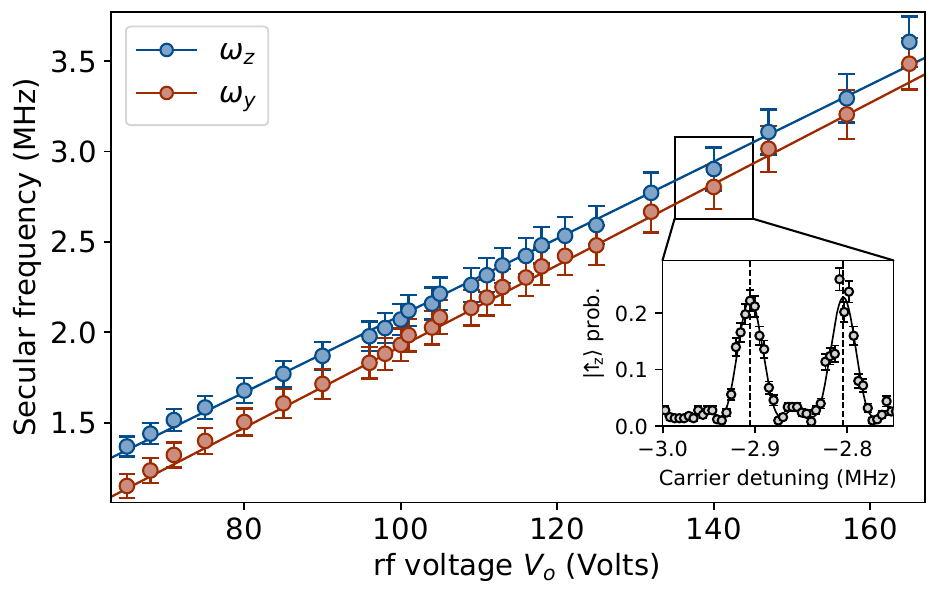}
    \caption{The radial secular frequencies in the $y$ and $z$ directions show an approximately linear dependence on the applied rf voltage. Solid lines are fits of Eq.~\ref{eq:trapfreqs} to the measured data. The inset shows a typical Raman-driven red sideband scan, from which the radial secular frequencies are extracted.}
    \label{fig:rfvssecular}
\end{figure}

We characterize our radial trap frequencies $\omega_y$ and $\omega_z$ as a function of applied rf voltage $V_0$ by probing the secular motional sidebands. Sidebands are driven using stimulated Raman transitions near 355 nm \cite{campbell2010ultrafast}, which typically couple to both radial modes. The rf voltage is coupled into the trap using a helical resonator \cite{siverns2012application} with a Q-factor of 275 at frequency $\Omega_t = 2\pi\times 28.543$~MHz. The resonator contains a $1\pm0.1\%$ capacitive pickoff to allow for monitoring of the rf voltage applied to the trap electrodes. In Fig. \ref{fig:rfvssecular}, we show the measured dependence of the radial trap frequencies on the applied voltage. The data are well described by the functional form of Eq.~\ref{eq:trapfreqs} and show approximately linear dependence on $V_0$. Given the small feature sizes of this trap design, we obtain trap frequencies in excess of $2\pi\times 3.5$~MHz with an applied rf voltage of only $V_0 = 165$~V.  

We also characterize the micromotion in this trap, which arises whenever ions are displaced from the rf null. In our trap geometry, we are particularly interested in micromotion along the $z-$direction, which coincides with the radial modes used for quantum simulation experiments in both 1D and 2D ion arrays. Using stimulated Raman beams at 355 nm, we drive carrier transitions at frequency $\omega_\text{hf}$ and the red micromotion sideband at frequency $\omega_\text{hf}-\Omega_t$. We then extract the corresponding Rabi frequencies $\Omega_0$ and $\Omega_{-1}$. The ratio of these frequencies $\Omega_{-1}/\Omega_0=\beta/2$ determines the micromotion modulation index $\beta=\Delta\vec{k}\cdot\vec{v_0}/\Omega_t$, where $\Delta\vec{k}$ is the wavevector difference of the Raman beams and $v_0$ is the ion velocity under driven micromotion \cite{keller2015precise,wang2020coherently}. 

\begin{figure}[b]
    \centering
    \includegraphics[width=\columnwidth]{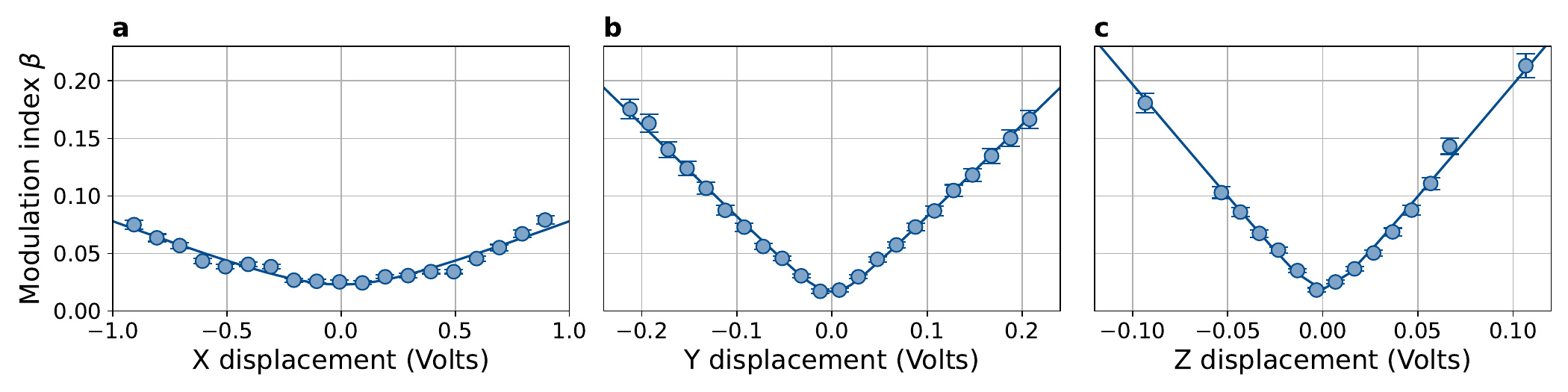}
    \caption{Micromotion modulation index $\beta$ measured along the $z$ direction, as a function of displacement voltage applied along the (a) x, (b) y, and (c) z trap axes. Zero displacement corresponds to the trap voltages that minimize the observed micromotion. Displacements are applied by asymmetrically biasing electrodes on opposite sides of the trap. Solid lines are fits to the data that include contributions from both intrinsic and extrinsic micromotion.}
    \label{fig:micromotion}
\end{figure}

In Fig. \ref{fig:micromotion}, we measure the micromotion modulation index $\beta$ as a function of voltage offsets in all three trap directions. Near the rf null, we measure a residual $\beta \lesssim 0.02$ which is attributed to the intrinsic micromotion (IMM) of the ion. Away from the null, the signal grows approximately linearly due to the contributions of the extrinsic micromotion (EMM). The fitted functions in Fig. \ref{fig:micromotion} account for both effects by setting $\Omega_{-1} = \sqrt{|\Omega_{-1,\text{IMM}}|^2+|\Omega_{-1,\text{EMM}}|^2}$. We observe that in the $y$ and $z$ directions of the trap, small voltage changes lead to measurable differences in micromotion; the full horizontal range of Fig. \ref{fig:micromotion}(b), for instance, corresponds to approximately $150~$nm real-space distance. In contrast, changes along the axial ($x$) direction are less sensitive to excess micromotion; Fig. \ref{fig:micromotion}(a) corresponds to approximately $40~\mu$m of total ion displacement.

\begin{figure}[h]
    \centering
    \includegraphics[width=\columnwidth]{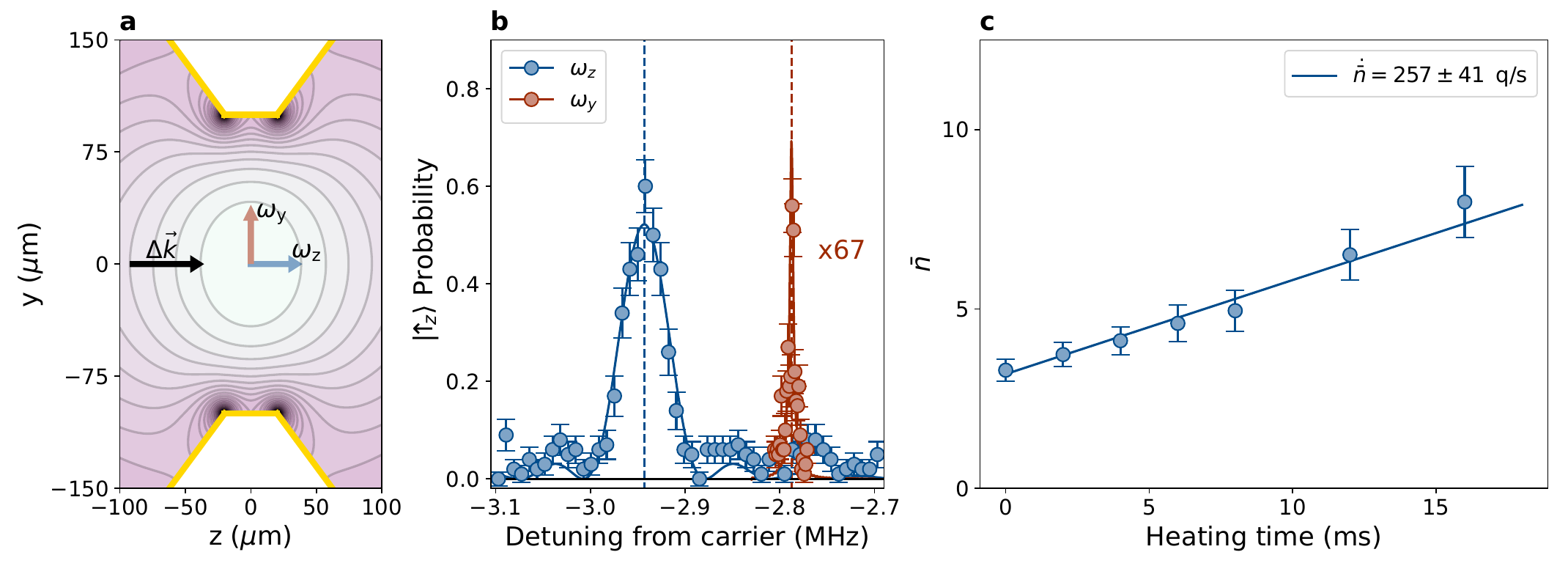}
    \caption{(a) Simulated trap potentials when the radial principal axes have been rotated to align with the trap coordinate axes. Under these conditions, the wavevector difference $\Delta \vec{k}$ of our Raman beams couples only to the $\omega_z$ mode. (b) When the principal axes are rotated, the $\omega_z$ radial mode (blue points) is excited during a Raman red-sideband scan. The $\omega_y$ mode (red points) is strongly suppressed and requires a 67 times longer Raman pulse duration to achieve similar excitation. (c) When the principal axes are rotated, we measure a center-of-mass heating rate of $257\pm 41$ quanta/s for the $\omega_z$ mode.}
    \label{fig:heatingrate}
\end{figure}

Rotating the principal axes of the trap is a critical capability, particularly when seeking to trap 2D lateral crystals. In the 2D lateral configuration, many ions experience micromotion since they are displaced from the rf null. Any projection of the micromotion velocity onto the $\Delta \vec{k}$ direction of the Raman beams therefore leads to unwanted rf heating of the modes used for quantum simulation. To suppress this effect, we tune our endcap, midcap, and center dc voltages to rotate the principal axes into alignment with the physical electrode axes (Fig.~\ref{fig:heatingrate}(a)). We experimentally confirm this rotation by probing the red-detuned radial secular sidebands and observing a near-total suppression of the radial $y$-mode. In Fig.~\ref{fig:heatingrate}(b), for instance, the $y-$mode is only visible when the Raman pulse duration is increased by a factor of 67.

Having rotated the principal axes to the experimentally desired orientation, we measure the center-of-mass heating rate along the $z$ direction. We begin by Doppler cooling a trapped ion to an average motional occupation $\bar{n}\approx4$ using red-detuned laser light at 369 nm. The ion is then allowed to heat for a variable delay time, after which $\bar{n}$ is determined by fitting Rabi oscillations on the red sideband. As shown in Fig.~\ref{fig:heatingrate}(c), the motional occupation increases linearly in time, corresponding to a heating rate of $\dot{\bar{n}}=257\pm 41~$quanta/s. This rate is less than half that of comparable laser-machined traps operating at room temperature \cite{wang2020coherently,menon2026monolithic}, despite the smaller ion-electrode distance in our trap. Further reductions in the heating rate may be achievable through ex-situ plasma cleaning, which has reduced heating rates by two orders of magnitude in similar laser-machined fused-silica traps \cite{menon2026monolithic}.

\subsection*{Ion Trapping and Manipulation}
\begin{figure}[t]
    \centering
    \includegraphics[width=\columnwidth]{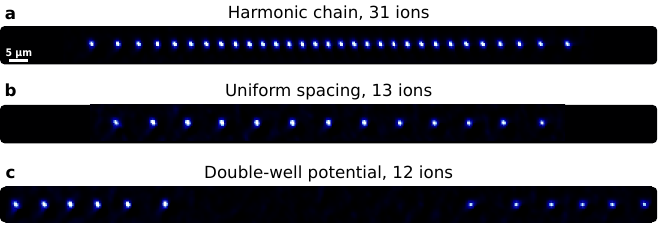}
    \caption{Ions trapped in different 1D configurations. (a) A 31-ion chain is trapped by keeping the axial confinement weak relative to the radial confinement. (b) Increasing the center electrode voltage produces a chain with near-uniform spacing. (c) Continuing to increase the center electrode voltage generates a symmetric double-well potential, with weak ion-ion interactions across wells.}
    \label{fig:1Dions}
\end{figure}

We further benchmark our design by trapping and cooling ions in a variety of different crystal configurations. To begin, we investigate the loading and stabilization of long 1D chains. Guided by our FEM simulations, we choose an initial configuration of voltages $\{V_{rf},V_{end},V_{mid},V_{ctr}\}=\{225,1.5,0,0\}$~Volts. This results in radial secular frequencies of $\sim 2\pi\times 2.5$~MHz and an axial frequency of $2\pi\times 180$~kHz. The relatively low axial frequency is required for long chains to prevent a transition from the 1D to zig-zag structural phase \cite{dubin1993theory}. Fig.~\ref{fig:1Dions}(a) shows a 31-ion 1D crystal trapped using these parameters, which remains stable under continuous Doppler cooling. We observe that compensation voltages of several volts are necessary to keep the chain centered; we attribute this to stray charges induced by our high-power 355 nm Raman beams and to micron-level misalignments of the trap substrates. 

The ion chain shown in Fig.~\ref{fig:1Dions}(a) exhibits a non-uniform spacing, which is typical for chains trapped in harmonic and near-harmonic potentials. However, many quantum simulation experiments may benefit from uniformly spaced ions \cite{johanning2016isospaced}, which facilitate individual ion addressing and multi-channel photomultiplier tube (PMT) readout. In addition, uniformly spaced ions lead to near-sinusoidal normal mode amplitudes and provide enhanced capabilities for engineering interactions between trapped ion qubits \cite{kyprianidis2024interaction}. In Fig.~\ref{fig:1Dions}(b), we show a linear chain of 13 ions with near-uniform spacing. For this configuration, we choose a set of trap voltages $\{V_{rf},V_{end},V_{mid},V_{ctr}\}=\{176,4.425,-4,1.88\}$~Volts to add anharmonicities to the axial potential and to equally balance the Coulomb repulsion with the trap confinement at each ion. We quantify the uniformity of the chain by comparing the standard deviation of ion spacings to the mean value, $\sigma_{\Delta x}/\overline{\Delta x}$ \cite{pagano2019cryogenic}. We find a spacing variation of $\sim4\%$, which is consistent with the theoretical minimum for a trap with 10 static electrodes. 

When the voltage applied to the central electrode is increased even further, it becomes energetically unfavorable for ions to remain near the center of the trap. Due to the symmetry of the trapping potential, the ions equilibrate between wells, as shown in Fig. \ref{fig:1Dions}(c). Such double-well systems have applications in many-body physics and in quantum chemical dynamics, where there are strong local interactions but only weak interactions across subsystems \cite{kyprianidis2024interaction}. Our split-well configuration, achieved with the voltage set $\{V_{rf},V_{end},V_{mid},V_{ctr}\}=\{176,4.53,-4,2.2\}$~Volts, has tunable separation distance and subsystem interaction strength based upon the voltage applied to the center electrode. We estimate a $20~$Hz inter-well coupling strength for the data shown in Fig. \ref{fig:1Dions}(c), which is $\approx5\%$ of the intra-well coupling between nearby ions. 

A core feature of our trap design is the ability to confine ions in both 1D chains and 2D arrays. In an rf trap, there are two possible orientations for a 2D ion crystal: a lateral 2D crystal, where the ion plane contains the axial direction \cite{wang2020coherently,kiesenhofer2023controlling,guo2024site}, and a radial 2D crystal, where the ion plane is perpendicular to the trap axis \cite{yoshimura2015creation,ivory2020paul,d2021radial,xie2021open}. In our system, we pursue the lateral 2D configuration so that ions can be imaged and addressed through the same viewport as for our 1D chains, which is not possible for a radial 2D crystal. We tune the axial frequency such that it becomes comparable to the $\omega_y$ radial frequency, resulting in a 2D crystal in the $xy$-plane. Our confinement along the $z$ direction remains strongest of all, to prevent the crystal from buckling into three dimensions.

\begin{figure}[t]
    \centering
    \includegraphics[width=.65\columnwidth]{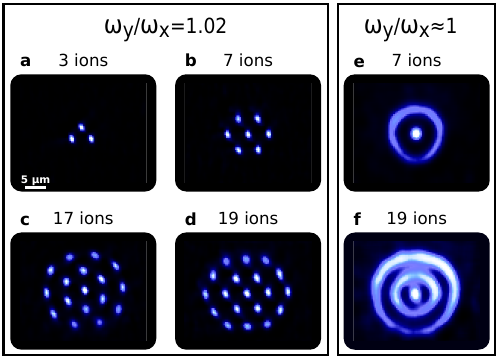}
    \caption{Ions trapped in near-circular 2D arrays. Crystals of (a) 3 ions, (b) 7 ions, (c) 17 ions, and (d) 19 ions self-assemble into a minimum-energy configuration when a slight frequency difference breaks the rotational symmetry of the trapping potential. When the potential symmetry is restored (e-f), the crystals are observed to rotate rapidly.}
    \label{fig:2dCircular}
\end{figure}

When the axial trap frequency is comparable to the $\omega_y$ radial frequency, the equilibrium configuration of ions is a nearly circular 2D crystal. In Fig.~\ref{fig:2dCircular}(a-d), we set $\omega_y \approx 2\pi\times 1.7$~MHz with $\omega_y/\omega_x = 1.02$ and observe a variety of 2D crystals containing $N=3-19$ ions. This small frequency asymmetry is introduced between the two in-plane directions to allow for stable trapping and prevent rotations of the 2D array. Though the configurations shown in Fig. \ref{fig:2dCircular} are stable for minutes under Doppler cooling, slight changes in the crystal energy or trap frequencies may induce rapid ion rearrangements or melting \cite{kiesenhofer2023controlling,pashinsky2025structural}. In Fig.~\ref{fig:2dCircular}(e-f), for example, we observe rotations of the trapped ion crystal when $\omega_y$ and $\omega_x$ are made approximately degenerate.

For larger 2D ion lattices, there are several reasons to avoid the near-degenerate case and impose an asymmetry between the axial and radial trap frequencies. First, larger asymmetries raise the energy barrier to spontaneous crystal rotation and avoid low-frequency vibrational modes that are more difficult to cool. Second, keeping $\omega_x < \omega_y$ reduces the high voltage and feature size demands of the endcap electrodes when shaping the axial potential. Third, the micromotion amplitude experienced by ions is proportional to their distance away from the trap axis. More ions can be trapped closer to the rf null when $\omega_x < \omega_y$, reducing exposure to large micromotion amplitudes and rf-driven heating.

\begin{figure}[h]
    \centering
    \includegraphics[width=\columnwidth]{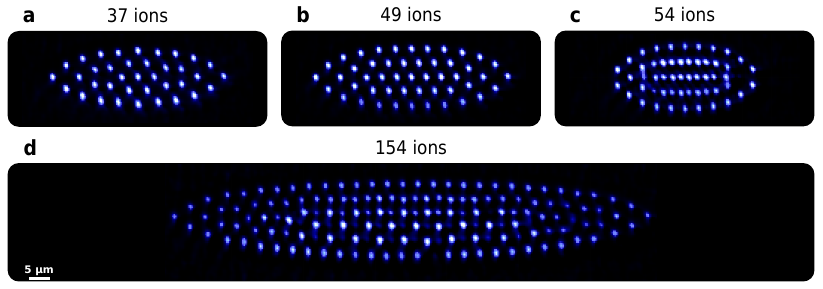}
    \caption{Ions trapped in lateral 2D arrays with large aspect ratio. Crystals of (a) 37 ions, (b) 49 ions, (c) 54 ions, and (d) 154 ions self-assemble into a variety of different lattice structures that depend sensitively on ion number and trap frequencies. Crystals in panels (a-c) are observed to be most stable given their mirror symmetry across both 2D axes.}
    \label{fig:2dPancake}
\end{figure}

In Fig.~\ref{fig:2dPancake}, we demonstrate the trapping of up to $N=154$ ions in large 2D arrays. We observe that the equilibrium lattice structure of these crystals depends sensitively upon ion number and trap voltages. For example, an $N=49$ ion crystal (Fig.~\ref{fig:2dPancake}(b)) with trap voltages $\{V_{rf},V_{end},V_{mid},V_{ctr}\}=\{152,15,3,5\}$~Volts leads to a 2D triangular lattice with no defects or dislocations. In contrast, an $N=37$ ion crystal under the same voltages (Fig.~\ref{fig:2dPancake}(a)) results in a mostly triangular lattice with symmetric dislocations in the bulk. Notably, an $N=54$ ion crystal under these conditions exhibits a rectangular lattice in the interior (Fig.~\ref{fig:2dPancake}(c)), which has not been previously predicted in the literature. For our largest crystals (Fig.~\ref{fig:2dPancake}(d)), we observe reduced crystal stability as ions frequently rearrange themselves between lattice sites. This may result from the relatively high pressure of $4\times 10^{-10}~$Torr within our vacuum chamber, which leads to frequent collisions between ions and background gas molecules. Additionally, unlike the crystals shown in Fig.~\ref{fig:2dPancake}(a-c), the one in Fig.~\ref{fig:2dPancake}(d) lacks mirror symmetry across both 2D axes; this reduced symmetry has been observed to increase susceptibility to ion rearrangements \cite{kiesenhofer2023controlling}.

\section*{Discussion}
In this work, we have introduced an ion trap for quantum simulation experiments in 1D and 2D arrays, at low fabrication complexity and with relative ease of assembly. Our design leverages commercially available selective laser etching technology to produce electrode geometries with small feature sizes and tight tolerances, which result in flexible axial potential shaping and high radial secular trap frequencies. Although the electrodes in our trap were fabricated from fused silica, our design is readily adaptable to other laser-machinable substrates such as alumina, sapphire, and diamond.

In our trap, we have demonstrated stable confinement in various 1D geometries, such as long chains, chains with uniform ion spacing, and ions split across two wells. Using only 2 Watts of input rf power, we have measured radial secular frequencies of $2\pi\times 3.5~$MHz for $^{171}$Yb$^+$ ions, which provides robust radial confinement for different axial potential configurations. We have also shown that micromotion can be detected via Raman-driven sidebands and compensated using small offset voltages applied to the trap electrodes. Finally, we rotate the trap principal axes to suppress Raman coupling to secular- and micro-motion along the $y$ direction, which is key for maintaining cold lateral 2D crystals. In this configuration, we measure a 250 quanta/s heating rate for the center-of-mass mode along the $z$ direction.

In addition, we have demonstrated the trapping of 2D arrays with up to $N=154$ ions. We believe this ion number to be limited by the relatively high background pressure in our vacuum chamber, which leads to frequent ion rearrangements due to collisions with background gases. Reducing the vacuum pressure by another 1-2 orders of magnitude using standard UHV techniques, or further still using a cryogenic ion trap \cite{pagano2019cryogenic}, is likely required to achieve stable arrays of $N > 200$ ions. We ultimately estimate that nearly $N\approx 1,000$ ions may be confined in the 2D geometry demonstrated here, before the micromotion amplitude for off-axis ions becomes comparable to the $\sim 5~\mu$m inter-ion spacing. However, it is likely that realizing these very large ion numbers in practice will require cryogenic-level vacuum pressures, highly symmetric ion lattices, and minimal cross-coupling between the axial and radial crystal modes.

\section*{Methods}

\subsection*{Trap Fabrication and Assembly}
The ability to host large 2D ion crystals and manipulate 1D geometries as shown above requires careful specification of the electrode geometry and applied trapping potentials. In our trap, we leverage selective laser etching (SLE) to define electrode features with 100-$ \mu$m-scale dimensions and 1-3 $\mu$m tolerances, verified with optical microscopy. SLE is typically performed in a two stage process. First, a high-power femtosecond laser is focused onto the substrate, locally altering the fused-silica structure. Second, the substrate is placed into an etching solution, dissolving the locally altered regions faster than the bulk \cite{bellouard2009SLE}. For materials that transmit wavelengths of light across a large range, complex internal features can also be created by focusing the laser into the substrate followed by chemical etching \cite{kiesenhofer2023controlling,menon2026monolithic}.

In this work, the trap substrates are selectively laser etched from fused silica ($\text{a}-\text{SiO}_{2}$) and fabricated by Translume. SLE of fused silica is a mature process, and fused silica exhibits a variety of material properties conducive to ion trapping such as a near zero thermal expansion coefficient, low charging in the presence of UV light, and a low dielectric constant. However, in comparison with other widely used substrates such as alumina, sapphire, and diamond, fused silica has a relatively low Young's modulus, requiring careful mechanical design to ensure the substrates can withstand the stresses of assembly. Moreover, fused silica has a comparatively low thermal conductivity, making it more susceptible to heating due to the application of high rf voltages.  

Following SLE, the substrates are metallized to create conductive trap electrodes and prevent line-of-sight dielectric exposure to the ions. SLE-fabricated deposition masks are placed onto the substrate to define the conductive regions of the trap. Metallization is performed by first depositing a thin adhesion layer of sputtered titanium, followed by a $1~\mu$m thick layer of sputtered gold. During deposition, substrates are rotated to ensure a uniform coating and full coverage between the gaps separating neighboring electrodes.

The alignment system for our trap substrates has been engineered to accommodate thermal expansion during high-temperature vacuum bakes while maintaining $1-3~\mu$m relative alignment tolerance between layers. Each substrate has been machined with four $1.000$~mm alignment holes that accept $1.000^{+ 0.000}_{- 0.014}$~mm dowel pins. The assembly is then clamped between Macor mounts with $1.181$~mm alignment holes. These clearances have been chosen to allow for thermal expansion of the components during a $\sim 2$ week $200^\circ$~C bake needed to achieve UHV pressures. Tight clearances between the dowel pins and fused silica substrates keep the electrodes well aligned, while looser clearances between the dowel pins and Macor reduce mechanical constrictions during the bake. 

In preliminary testing of the substrate assembly, the Macor mounting system was manufactured with $1.008~$mm alignment holes. We found that this small clearance between the hole and dowel pins was not sufficient to accommodate the differential thermal expansion between fused silica and Macor during baking, resulting in cracking along the alignment holes in the substrates. Although this cracking occurred away from the electrode traces and the substrates remained mechanically clamped, the long term reliability following multiple bake cycles could not be assured. As a result, we enlarged the Macor alignment holes to $1.181~$mm to allow for thermal expansion of the Macor without transferring stresses to the fused silica.

Delivering voltages to the trap electrodes also posed a challenge due to the fragility of the fused silica substrates and the close proximity of the dc and rf layers. For this reason, we mounted two custom alumina PCBs (CERcuits) as interposers between the substrates and the connections to vacuum electrical feedthroughs. These PCBs are double-sided and gold-plated, with metallized vias to simplify the electrical routing. The PCBs and dc substrates are connected using $12.7~\mu$m~$\times$~$508~\mu$m gold ribbons (California Fine Wire) and secured into place with UHV conductive epoxy (EPO-TEK H20E-PFC) (Fig.~\ref{fig:trapcad}(b)). On the PCBs, $1000~$pF capacitors connect each independent dc electrode to ground, shunting rf pickup. The PCB grounds are connected to the mounting screws on the primary trap frame, providing a common ground for the dc power supply, vacuum chamber, rf resonator, and grounded dc electrodes.

The rf voltage is delivered by applying up to 2 W of power at 28 MHz to a single-coil helical resonator \cite{siverns2012application}. A capacitive divider at the output of the resonator allows for probing and stabilization of the rf voltage at the trap \cite{johnson2016rfstabilization}. Inside the vacuum chamber, rf voltage is delivered to the trap substrate via mechanical connection. Gold ribbons are epoxied to the rf electrode and crimped to a $\#2-56$ lug connector, which is attached to the rf feedthrough wire.
 
The trap is housed in a Magdeburg hemisphere vacuum chamber (Kimball Physics MCF450-MH10204/8-A). A front $4.5~$" conflat viewport provides optical access to the trap for imaging and state manipulation, and two $2.75~$" conflat side viewports at $45^\circ$ provide additional laser access. Inside the chamber, two groove grabbers serve as the main attachment point for a primary aluminum frame, onto which all trap components are mounted. The electrode assembly is fastened to a secondary aluminum frame so that its delicate alignment may be performed outside the vacuum chamber, then secured to the primary frame afterwards. Three $1.33~$" conflats are used for electrical connections to the dc and rf electrodes as well as to the atomic ovens. A fourth $1.33~$" conflat provides the vacuum connection to the ion and non-evaporable getter pumps. 

\subsection*{Ion Trapping, Cooling, Detection, and Manipulation}
In this work we use $^{171}$Yb$^+$, which exhibits a long-lived hyperfine ground state qubit encoded in the $^{2}\text{S}_{1/2}|F=0,m_{F}=0\rangle\equiv\ket{\!\downarrow}$ and $^{2}\text{S}_{1/2}|F=1,m_{F}=0\rangle\equiv\ket{\!\uparrow}$ states. The qubit levels are separated by $12.6~$GHz, and the degeneracy of the $F=1$ states is broken by a 5 G magnetic field that is oriented perpendicular to one of the 2.75" conflat viewports. This magnetic field orientation was chosen to facilitate the use of $\sigma^{\pm}$-polarized beams for future Electromagnetically-induced-transparency cooling. Cooling, state preparation, and detection are performed by illuminating the ions with near- or on-resonant 369.5 nm light \cite{olmschenk2007manipulation}. Ion fluorescence is captured by a 0.27 NA lens and directed to either a photomultiplier tube (Hammamatsu H10682-210) or electron-multiplied CCD camera (Andor iXon 897). For Figures \ref{fig:1Dions}, \ref{fig:2dCircular}, and \ref{fig:2dPancake}, ion images were captured with the camera and post-processed to correct for optical aberrations.

The $^{171}$Yb$^+$ ions are produced by photoionizing isotopically-enriched neutral Yb emitted from a thermal source. Although our trap typically operates with secular frequencies above 2 MHz, we observe that loading large ion crystals at these frequencies is inefficient. For the crystals in Figures \ref{fig:1Dions}, \ref{fig:2dCircular}, and \ref{fig:2dPancake}, we found that a highly anharmonic potential generated by the voltage set $\{V_{rf},V_{end},V_{mid},V_{ctr}\}=\{150,0.75,0,0.55\}$~Volts provided fast and reliable loading for large ion crystals. The large anharmonicity increases the trapping volume, resulting in lower Coulomb interaction energies between ions and more efficient laser cooling. For crystals exceeding $\sim 50$ ions, the lowest energy configuration for these voltages is in the zig-zag phase, which remains stable while additional ions are loaded. Once the desired number of ions has been loaded with this voltage set, we ramp the trap voltages to their final values and allow ions to self-assemble into their equilibrium positions.

Coherent, counter-propagating Raman beams at 355 nm are used to drive coherent quantum operations and spin-motion coupling. The beams are derived from a mode-locked laser with 15 ps pulses at a repetition rate of 80 MHz. Each Raman beam passes through an acousto-optic modulator to enable precise control over the frequency difference between beams. The wavevector difference of the Raman beams, $\Delta \vec{k}$, is oriented perpendicular to the trap axis and couples only to ion motion along the trap $z$ direction. When the trap principal axes have not been rotated (Fig.~\ref{fig:rfvssecular}), Raman transitions can drive both radial modes since both have projections along the trap $z$ direction. Following rotation of the principal axes (Fig.~\ref{fig:heatingrate}(b)), only the mode co-aligned with the $z$ axis is visible. Throughout this work, Raman transitions have also been used to probe the micromotion component along the $z$ direction (Fig.~\ref{fig:micromotion}) and to drive red sideband transitions for ion heating rate measurements (Fig.~\ref{fig:heatingrate}(c)).

\section*{Data Availability Statement}
The datasets and analysis files generated during the current study are publicly available through the Indiana University DataCore repository. Also available in this repository are the full CAD models, parts lists, and assembly instructions for the ion trap described in this work.

\bibliography{sample}

\section*{Acknowledgments}

This work was supported by the Gordon and Betty Moore Foundation, grant DOI 10.37807/GBMF12963 and by the National Science Foundation under Grant No. PHY-2412878.

\section*{Author contributions statement}

F.G.S. designed and fabricated the ion trap discussed in this work. F.G.S. and P.R. conceived the experiments. F.G.S., I.J., T.W.B., J.L., J.W.V.V., and J.E. built the surrounding experimental hardware and conducted the experiments. F.G.S., I.J., T.W.B, and J.L. analyzed the data. All authors participated in the writing and editing of the manuscript.

\section*{Competing Interests}

The authors declare no competing interests.

\end{document}